# Relaxation and pumping of quantum oscillator nonresonantly coupled with the other oscillator

A.I.Trubilko
St. Petersburg State University of State Fire Service of Emercom of Russia,
St. Petersburg, 196105 Russia
e-mail: trubilko.andrey@gmail.com

A.M.Basharov
National Research Center "Kurchatov Institute,"
Moscow, 123182 Russia
Moscow Institute of Physics and Technology (Technical University),
Dolgoprudnyi, Moscow oblast, 141701 Russia
e-mail: basharov@gmail.com

The paper shows mechanisms of both the pumping and energy decay of an "isolated" oscillator. The oscillator is only non-resonantly coupled with the adjacent oscillator which resonantly interacts with the thermal bath environment. Under these conditions the "isolated" oscillator begins interacting with the thermal bath environment of the adjacent oscillator. The conclusion is based on the kinetic equation derived relative to anti-rotating terms of the initial Hamiltonian, with the latter being the Hamiltonian of two oscillators and environment of one of them.

Quantum oscillators simulate photon systems in micro resonators, which can be either coupled with each other on mirrors or with pump and vacuum (thermal bath) fields of the thermal bath, providing an example of an open quantum system. Interaction with thermal baths leads to the attenuation of quantum oscillators. Effective research of attenuating quantum systems were made possible due to the introduction of the notion of a kinetic equation (master equation) into the mathematical apparatus of nonlinear and quantum optics [1-3].In [4-5], the general form of the kinetic equation is established, that is now commonly referred to as the Lindblad-type of the kinetic equation. Many recently published papers on the analysis of open quantum systems dynamics start with the definition (as initial ones) of precisely kinetic equations in the Lindblad form with a predefined Lindblad operators.

In view of this, researchers consider atomic systems interacting with electromagnetic fields of various nature [6–9], photon systems consisting of photons of cavity modes interacting with other cavity systems, with intracavity and boundary atoms [10–13], and other optical problems.

Since the relaxation channels are already defined by the appropriate terms and Lindblad operators in the initial equations, the majority of further approximations, and, in particular, dispersion approximation do not adequately describe the case under study. In fact, in optics, the initial Hamiltonian of two non-resonantly interacting atoms has both rapidly and slowly varying terms. This is distinctly seen from the Hamiltonians of the basic models of quantum optics in the interaction representation [14, 15]. The rapidly varying terms in the interaction representation are commonly referred to as anti-rotating ones. It was observed that in optical problems the success of the approach based on kinetic equations in the Lindblad form is due to the neglect of anti-rotating terms [16]. However, the neglect is possible only in resonant processes, and not always in all cases [17, 18]. Thus it is vital to take into account the anti-rotating terms in deriving the kinetic equation for resonant, quasi-resonant and non-resonant processes, as well as to analyze their significance in optical effects.

The present paper considers the dynamics of a quantum oscillator that is non-resonantly coupled to another oscillator. The connection of the kind is often neglected, assuming that the given oscillator can be considered to be isolated from the other one. Our work has shown that the non-resonant coupling provides both the pumping of the oscillator and its decay due to the pumping and relaxation channels of the oscillator whose non-resonant coupling is generally neglected. In addition, it is demonstrated that an oscillator interacting indirectly with a bath forms its own relaxation channel in the absence of any resonant interactions with the medium. For example, if there are two quantum oscillators of significantly different frequencies $\omega_c \neq \omega_r$, with the oscillator $\omega_r$ being "virtually isolated" and interacting merely non-resonantly with the oscillator $\omega_c$, then this non-resonant interaction forms a direct relaxation channel of the oscillator $\omega_r$ if the oscillator $\omega_c$ interacts with a thermal bath field. In this case, the oscillator $\omega_c$ interacts with its region of the spectrum of the thermal bath field, whose central frequency is equal to the frequency $\omega_c$, i.e. it is resonant to frequency of the oscillator $\omega_c$. The oscillator $\omega_r$ that is uncoupled with a thermal bath interacts with the same bath as the oscillator $\omega_c$, only with thermal bosons, whose frequencies lie in another region of the spectrum, namely $\omega_r$. It can be clearly seen in the figure below.

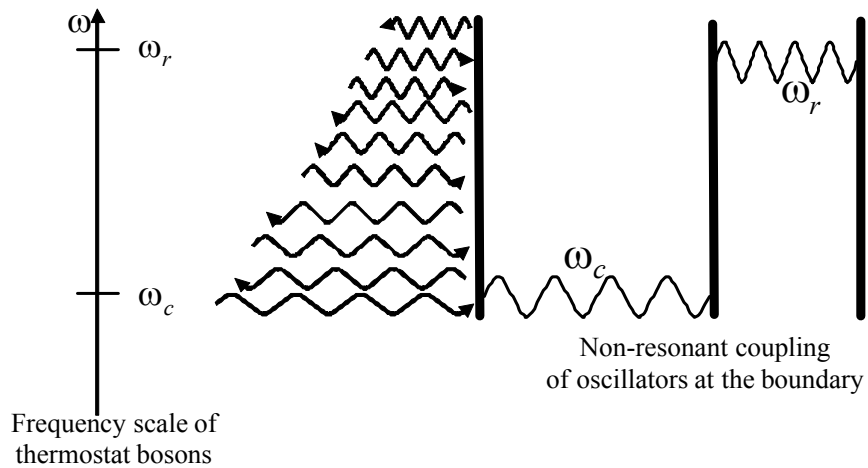

Figure.

The figure caption. Oscillators $\omega_c$ and $\omega_r$ are presented as modes of two non-resonant coupled cavities. As a consequence of transforming the system's initial Hamiltonian by means of the algebraic perturbation theory, the resulting effective Hamiltonian describes the interaction of cavity modes with different regions of the spectrum of thermal bosons whose central frequencies coincide with frequencies $\omega_c$ and $\omega_r$.

It is to be emphasized that basing on the kinetic equation with predefined relaxation operators in the form of Lindblad due to neglect of the anti-rotating terms, all oscillators should interact only with the spectral region of thermal bosons with the central frequency $\omega_c$. However, in fact, thermal bath fields can be modeled by high-intensity noise sources, whose frequencies are scattered around certain central frequencies, and therefore may not overlap at all and are characterized by different parameters, for example, the number density of photons per unit value of the frequency spectrum. These sources have been considered as independent noise sources starting from [19]. It means that such processes cannot be described by the dispersion limit of the kinetic equation with the predefined Lindblad operators, and a new kinetic equation should be derived according to the posed problem.

It is to be noted that is a similar problem arose in describing attenuating classical oscillators [20].

The case of "purely" photon (boson) systems is of particular importance in terms of kinetic equations. On the one hand, although there have been developed methods for the exact solution (the Hamiltonian diagonalization) of multiparticle and multimode boson problems [21, 22], they still need to be numerically simulated. That is the so-called global approach in thermodynamic problems. On the other hand, in accordance with [21], the resonance approximation for interacting modes gives rise both to the results that contradict thermodynamics principles (energy transfer from the cold to the hot environment is made "possible") and an incorrect stationary state of strongly coupled quantum systems [23]. The latter was first noted in [24] by introducing the phenomenological relaxation operator for resonantly interacting photon systems.

That being the case, it is undoubtedly important to derive the kinetic controlling equation of an open system from both general principles and the initial Hamiltonian under the conditions of non-resonant (dispersive) interaction of photon modes.

The present paper considers the case of two non-resonantly interacting oscillators with account of anti-rotating terms of their interaction operator using the algebraic perturbation theory. It also describes both the mechanism of forming the so-called interference [18] interactions by involving the interaction of one of the oscillators with a thermal bath field, and the relaxation channel with the thermal bath of an oscillator that is not directly coupled with the thermal bath. Finally, both a kinetic equation and its solution have been obtained, which take into account all anti-rotating terms and do not contradict thermodynamics principles, as evidenced by its standard Lindblad form, in terms of which two relaxation operators are presented for the predefined and new relaxation channels.

## 1. Non-resonantly coupled cavities

A quantum oscillator with the Hamiltonian $H_c = \hbar\omega_c c^+ c$ is the simplest quantum model. It successfully describes photons in high-quality single-mode cavities, plasmon oscillations and other nanoobjects; interactions of an oscillator with various objects - electromagnetic fields, atoms, other cavities, etc. - have long been given substantial consideration. Such an oscillator will be called - by its frequency - the oscillator $\omega_c$. The simplest case of the two interacting oscillators, for example, $\omega_c$ and $\omega_r$ is described by the Hamiltonian [24] $H_c + H_r + V_{c-r}$, where a general form of the interaction operator $V_{c-r}$ is defined by the interaction parameter (the coupling constant) $g$:

$$V_{c-r} = g(c + c^+)(r + r^+),$$

where pairs of annihilation and creation operators $r$, $r^+$ and $c$, $c^+$ of photons $\omega_r$ and $\omega_c$ satisfy the usual Bose commutation relations, and at the same time, the operators of each pair commute with the operators of the other pair.

The problem of the dynamics of two interacting oscillators was solved in a general form in work [25]. In the same place, inconsistencies with an approximate method were analyzed, when the anti-rotating terms were neglected. As a result, it is possible to neglect the anti-rotating terms only in the case of resonant interaction.

At the same time, the exact results looked rather cumbersome, becoming obscure with an additional account of interaction of one of the oscillators with the bosons of the thermal bath field, allowing for their further application only in numerical counting [22].

Meanwhile, treating a multimode field as a thermal bath suggests that the Markov approximation [21] be used, so that, generally speaking, precise results are not needed in this case. It is required that computations be clear to analyze possible physical consequences of the interaction of one of the oscillators with bosons of the thermal bath, which cannot be easily seen in the precise approach, and, for the time being, such results are unknown to the authors to solve the problem to be considered below. In the case of arbitrary frequencies, including also the limiting cases or the reverse, as well as the case of close frequencies, the main feature of optical systems is visible.

A major feature of optical systems is evident from the case of arbitrary frequencies $\omega_c \neq \omega_r$, involving limit $\omega_c \gg \omega_r$ or reverse ones $\omega_c \ll \omega_r$, as well as the case of close frequencies $\omega_c \approx \omega_r$. If one writes down the Schrödinger equation in the interaction representation

$$i\hbar \frac{d}{dt}|\Psi(t)\rangle = V_{c-r}(t)|\Psi(t)\rangle, \qquad (1)$$

$$V_{c-r}(t) = g(ce^{-i\omega_c t} + c^+ e^{i\omega_c t})(re^{-i\omega_r t} + r^+ e^{i\omega_r t}),$$

it can be seen that the terms have rapidly time-varying factors $e^{\pm i(\omega_c \pm \omega_r)t}$. In the case of close frequencies there occur slowly varying terms with factors $e^{\pm i(\omega_c - \omega_r)t}$ apart from rapidly time-varying ones.

A standard approach to simplifying equation (1) is to use the Krylov-Bogolyubov-Mitropolsky averaging method [26-28]. We will first demonstrate its application for the case of non-resonantly interacting oscillators. Then, by averaging, we immediately obtain that $\langle V_{c-r}(t) \rangle = 0$, so that non-resonant oscillators can be considered as non-interacting ones in the first approximation.

In order to take into account the second order of the averaging method as applied to similar optical problems, it is convenient to use its algebraic alternative [17, 18, 29]. Using the unitary symmetry of quantum theory, let us turn to a new representation with the help of the unitary transformation $|\widetilde{\Psi}(t)\rangle = \exp(-iS)|\Psi(t)\rangle$. In the new representation, all relations, including the Schrödinger equation, have a previous form but are marked with a "tilde" sign. Expanding (according to the general theory [18]) the transformed Hamiltonian and transformation generator in series over the coupling constant $g$, we obtain

$$S(t) = S^{(1)}(t) + S^{(2)}(t) + \ldots, \quad \widetilde{V}_{c-r}(t) = \widetilde{V}^{(1)}_{c-r}(t) + \widetilde{V}^{(2)}_{c-r}(t) + \ldots$$

where the Baker-Hausdorff formula is taken into account

$$\widetilde{V}^{(1)}_{c-r}(t) = \hbar \frac{dS^{(1)}(t)}{dt} + V_{c-r}(t), \qquad (2)$$

$$\widetilde{V}^{(2)}_{c-r}(t) = \hbar \frac{dS^{(2)}(t)}{dt} - \frac{i}{2}[S^{(1)}(t), \widetilde{V}^{(1)}_{c-r}(t)] - \frac{i}{2}[S^{(1)}(t), V_{c-r}(t)]. \qquad (3)$$

The major requirement that meets the Krylov-Bogolyubov-Mitropolsky approach is that there are no rapidly time-varying terms in the transformed Hamiltonian. Time averaging leads to equality $\widetilde{V}^{(1)}_{c-r}(t) = 0$, which also corresponds to our assumption about the absence of rapidly time-varying terms. However, we additionally obtain the value of the transformation generator. Under the assumption of the adiabatic inclusion of the interaction we have

$$S^{(1)}(t) = cr \frac{ge^{-i(\omega_c+\omega_r)t}}{i\hbar(\omega_c+\omega_r)} - c^+r^+ \frac{ge^{i(\omega_c+\omega_r)t}}{i\hbar(\omega_c+\omega_r)} + cr^+ \frac{ge^{-i(\omega_c-\omega_r)t}}{i\hbar(\omega_c-\omega_r)} - c^+r \frac{ge^{i(\omega_c-\omega_r)t}}{i\hbar(\omega_c-\omega_r)}.$$

According to the formula (3), this generator determines the second order correction in the coupling constant due to the account of the anti-rotating terms. The absence of the rapidly time–varying terms in (3) obeys the requirements of the Krylov-Bogolyubov-Mitropolsky approach:

$$\widetilde{V}^{(2)}_{c-r}(t) = -c^+ c \Pi_c(\omega_r) - r^+ r \Pi_r(\omega_c) - \frac{g^2}{\hbar(\omega_c+\omega_r)}, \qquad (4)$$

$$\Pi_c(\omega_r) = \frac{g^2}{\hbar}\left(\frac{1}{\omega_c+\omega_r}+\frac{1}{\omega_c-\omega_r}\right), \quad \Pi_r(\omega_c) = \frac{g^2}{\hbar}\left(\frac{1}{\omega_r+\omega_c}+\frac{1}{\omega_r-\omega_c}\right). \tag{5}$$

It can be seen that the oscillators remain non-interacting with each other in the second order, nevertheless, the impact of the other oscillator is apparent in the value of the parameters $\Pi_c(\omega_r)$ and $\Pi_r(\omega_c)$ which determine frequency shifts.

The case of resonant interaction is also described at $\omega_c \approx \omega_r$. It is to be emphasized that we are analysing non-resonantly interacting oscillators $\omega_c$ and $\omega_r$, when their frequencies differ significantly from each other. In case of resonantly interacting oscillators there occur divergent resonant denominators in expressions for $\Pi_c(\omega_r)$ and $\Pi_r(\omega_c)$. Divergent denominators should be excluded, but then, according to formula (1), the interaction operator becomes a nonzero operator $\widetilde{V}_{c-r}^{(1)}(t)$. As a result, the following effective interaction operators and a transformation generator are obtained for the resonant interaction

$$\widetilde{V}_{c-r}^{(1)}(t) = g(cr^+ e^{-i(\omega_c-\omega_r)t} + c^+ r e^{i(\omega_c-\omega_r)t}), \tag{6}$$

$$\widetilde{V}_{c-r}^{(2)}(t) = -c^+ c\Pi(\omega_c) - r^+ r\Pi(\omega_c) - \frac{g^2}{2\hbar\omega_c}, \quad \Pi(\omega_c) = \frac{g^2}{2\hbar\omega_c}, \tag{7}$$

$$S^{(1)}(t) = cr\frac{ge^{-i(\omega_c+\omega_r)t}}{i2\hbar\omega_c} - c^+ r^+ \frac{ge^{i(\omega_c+\omega_r)t}}{i2\hbar\omega_c}.$$

The transformation generators $S^{(1)}(t)$ determine not only the second order of the Hamiltonians of the non-resonantly interacting oscillators $\omega_c$ and $\omega_r$, but also the so-called interference channels with allowance for any other interactions [18].

## 2. The decay channel to the thermal bath of one of the non - resonantly coupled oscillators with the additional interaction of the other oscillator with the thermal bath

Now let one of the non-resonantly interacting oscillators, e.g., $\omega_c$, be coupled with a thermal bath. This is described by the Hamiltonian of the thermal bath $\sum_\omega \hbar\omega a_\omega^+ a_\omega$ and the interaction operator $V_c$ of the oscillator $\omega_c$ with a thermal bath, which can be written down with account of all anti-rolling terms in the interaction representation:

$$V_c(t) = \gamma_c \sum_\omega (ce^{-i\omega_c t} + c^+ e^{i\omega_c t})(a_\omega e^{-i\omega t} + a_\omega^+ e^{i\omega t}). \tag{8}$$

Here $\gamma_c$ is the coupling constant with a thermal bath, whose creation operators $a_\omega^+$ and annihilation operators $a_\omega$ obey ordinary Bose commutation relations. Account of the interaction in the problem of two non-resonantly interacting oscillators $\omega_c$ and $\omega_r$ consists in changing the expansion of generator $S$ of transformation of the system's wave vector and transformed total interaction operator $V(t) = V_c(t) + V_{c-r}(t)$:

$$S(t) = S^{(1,0)}(t) + S^{(0,1)}(t) + \ldots, \quad \widetilde{V}(t) = \widetilde{V}^{(1,0)}(t) + \widetilde{V}^{(0,1)}(t) + \widetilde{V}^{(1,1)}(t) + \ldots$$

Now, the left superscript corresponds to the nonresonance interaction between oscillators, which, as before, describes the order of $g$ constant. The right index corresponds to the interaction of the oscillator $\omega_c$ with a thermal bath and marks the order of the term of $\gamma_c$ constant. Interference terms are determined by the following expression of the algebraic perturbation theory

$$\widetilde{V}^{(1,1)}(t) = \hbar \frac{dS^{(1,1)}(t)}{dt} -$$

$$-\frac{i}{2}[S^{(1,0)}(t), V_c(t)] - \frac{i}{2}[S^{(1,0)}(t), \widetilde{V}^{(0,1)}(t)] - \frac{i}{2}[S^{(0,1)}(t), V_{c-r}(t)] - \frac{i}{2}[S^{(0,1)}(t), \widetilde{V}^{(1,0)}(t)],$$
(9)

The calculations described in the previous section have given rise to the effective Hamiltonian $V^{Eff}(t)$ of the problem of the two non-resonant interacting oscillators under conditions when the oscillator $\omega_c$ is additionally connected to the thermal bath $V^{Eff}(t)$

$$V^{Eff}(t) = \widetilde{V}_c^{(1)}(t) + \widetilde{V}_r^{(2)}(t) + \widetilde{V}_{c-r}^{(2)}(t),$$
(10)

$$\widetilde{V}_c^{(1)}(t) = \gamma_c \sum_{\omega \in (\omega_c)} (c a_\omega^+ e^{-i(\omega_c - \omega)t} + c^+ a_\omega e^{i(\omega_c - \omega)t}),$$

$$\widetilde{V}_r^{(2)}(t) = -\frac{g\gamma_c}{2\hbar\omega_c} \sum_{\omega \in (\omega_r)} r^+ a_\omega e^{-i(\omega - \omega_r)t} - \frac{g\gamma_c}{2\hbar\omega_c} \sum_{\omega \in (\omega_r)} r a_\omega^+ e^{i(\omega - \omega_r)t}.$$

The expression for $\widetilde{V}_{c-r}^{(2)}(t)$ is given by formulae (4) and (5). The operator $\widetilde{V}_c^{(1)}(t)$ effectively describes the interaction of the oscillator $\omega_c$ with the thermal bath introduced into the problem after averaging over the rapidly varying terms of the initial Hamiltonian (8). Interaction is more effective with those bosons, whose frequencies lie near the central resonant frequency $\omega_c$ (see the figure). This region of spectrum of thermostat bosons is designated as $(\omega_c)$. The operator $\widetilde{V}_r^{(2)}(t)$ describes resonant interaction of oscillator $\omega_r$ with the same thermal bath. This interaction has resulted from the interference of non-resonant interaction between oscillators $\omega_c$ and $\omega_r$ (their initial interaction operator is wholly defined by anti-rotating terms) and non-resonant anti-rotating terms of the operator (8) of interaction of the oscillator $\omega_c$ with the thermal bath. Interaction is more effective with those bosons, whose frequencies lie near the central resonant frequency $\omega_r$ (see the figure). This region of spectrum of thermostat bosons is designated as $(\omega_r)$.

Operators $\widetilde{V}_c^{(1)}(t)$ and $\widetilde{V}_r^{(2)}(t)$ have a standard form that allows them to be represented in the Markov approximation by quantum Wiener processes (see, for example, [16-18,30-32]) and to write down an equation for the evolution operator in the form of a stochastic differential equation (SDE). Further, the kinetic equation for the density matrix $\rho^S(t)$ of non-resonantly interacting oscillators $\omega_c$ and $\omega_r$ is obtained in the standard way (the upper index $S$ indicates the system of two non-resonantly coupled oscillators). Since the effective non-resonantly coupled oscillators did not turn out to interact with each other, and second order constant frequency shifts (7) can be included in the renormalized frequencies of oscillators $\widetilde{\omega}_c$ and $\widetilde{\omega}_r$, $\widetilde{\omega}_c = \omega_c - \Pi_c(\omega_r)$, $\widetilde{\omega}_r = \omega_r - \Pi_r(\omega_c)$, the kinetic equation has the most standard Lindblad form in the interaction representation

$$\frac{d\rho^S(\bar{t})}{d\bar{t}} = -\widehat{\Gamma}_c \rho^S(\bar{t}) - \widehat{\Gamma}_r \rho^S(\bar{t}),$$
(11)

$$\hat{\Gamma}_i \rho^S(\bar{t}) = -\bar{\gamma}_i \bar{n}_i Y_i^+ \rho^S(\bar{t}) Y_i - \bar{\gamma}_i Y_i \rho^S(\bar{t})(\bar{n}_i + 1) Y_i^+ +$$
$$+ \left( \bar{\gamma}_i (\frac{\bar{n}_i + 1}{2} Y_i^+ Y_i + \frac{\bar{n}_i}{2} Y_i Y_i^+) \rho^S(\bar{t}) + \rho^S(\bar{t}) \bar{\gamma}_i (\frac{\bar{n}_i + 1}{2} Y_i^+ Y_i + \frac{\bar{n}_i}{2} Y_i Y_i^+) \right).$$

Here the index $i$ numbers non-resonantly coupled oscillators of the open system $\omega_c$ and $\omega_r$, ranging over values $c$ and $r$; a line above the symbol indicates the dimensionless analogue of the value introduced earlier: $\bar{t} = \omega_c t$, $\bar{\gamma}_c = \dfrac{2\pi\gamma_c^2}{\hbar\omega_c^2}$, $\bar{\gamma}_r = \dfrac{\pi g^2 \gamma_c^2}{2\hbar^2 \omega_c^4}$. In the coupling constants with thermal baths, their transformation is taken into account when obtaining the kinetic equation [16-18,30-32]. The annihilation operators prove to be equal to Lindblad's operators $Y_c = c$, $Y_r = r$ with allowance for the renormalization of the coupling constants with the thermal bath. Finally, the thermodynamic parameters — densities of the number of photons $\bar{n}_c$ and $\bar{n}_r$ per unit dimensionless frequency — are determined at frequencies $\omega_c$ and $\omega_r$, respectively, i.e. if the average of creation and annihilation operators of bosons' thermal bath is delta-correlated $<a_\omega^+ a_{\omega'}> = n(\omega)\delta(\omega - \omega')$, then $\bar{n}_c = n(\omega_c)\omega_c^{-1}$, $\bar{n}_r = n(\omega_r)\omega_c^{-1}$. It is to be noted that these photon densities correspond to photon densities of intense chaotic boson fields, which can simulate various spectral regions of the boson delta-correlated field interacting with an oscillator $\omega_c$.

If the initial states of non-resonantly coupled oscillators are not entangled in any way, then equation (11) falls into two equations, with each of them describing one oscillator, resonantly coupled with a thermal bath field. Then the average number of the oscillator's photons in the stationary state $<Y_i^+ Y_i> = n_i$, so that there arises no contradiction with the thermodynamics laws in the suggested approach.

### 3. Conclusion

The suggested approach does not describe any phenomenological modeling of processes and phenomena, but is based exclusively both on the first principles and natural assumption that the open quantum system interacts with the thermal bath according to Markovity. In order to derive the kinetic equation (11), there is no need to use a complicated global approach and Hamiltonian diagonalization – as an alternative, the Hamiltonian of non-resonantly coupled systems decay into diagonal Hamiltonians of non-interacting oscillators while applying the Krylov-Bogolyubov-Mitropolsky averaging method. When applied to optical problems, this method is expressed in its algebraic version developed in [17,18,29], into which the method of quantum SDE is naturally integrated for the basic kinetic equation in the Markov approximation to be obtained [18]. Up to now, the cases of integration of quantum SDEs into algebraic perturbation theory have been considered only for open quantum systems with an atomic subsystem [36-38]. The algebraic version of the Krylov-Bogolyubov-Mitropolsky averaging method (otherwise, algebraic perturbation theory [29]) makes it possible to visually take into account all the anti-rotational terms of the initial boson Hamiltonian, whose interference is described by means of unitary transformation generators of the initial state vector of the entire system that consists of non-resonantly coupled oscillators and a thermal (delta-correlated) field interacting with one of the oscillators. As a result, a "non-interacting" oscillator, that is only non-resonantly coupled to the other one, forms a relaxation and/or pumping channel, exchanging energy with a thermal bath field and other external fields that are directly unrelated to it. It is obvious that such channels usually overlooked in the previous studies can be found in other quantum systems except for optical ones.